\documentclass{article}

\usepackage[final,sglblindworkshop]{neurips_2026}
\workshoptitle{First Workshop on Interpreting Agent Behavior (IAB) Competition Paper Track}

\usepackage[utf8]{inputenc}
\usepackage[T1]{fontenc}
\usepackage{amsmath,amsfonts}
\usepackage{booktabs}
\usepackage{graphicx}
\usepackage{float}
\usepackage{microtype}
\usepackage{url}
\usepackage{hyperref}
\hypersetup{
  hidelinks,
  pdftitle={How Far Can an Agent Push? Trust-Safety Margins in Live Language-Based Economic Games},
  pdfauthor={Dima Galat and Marian-Andrei Rizoiu},
  pdfsubject={Interpreting Agent Behavior (IAB), NeurIPS 2026 Competition Paper Track}
}

\newcommand{\qstar}{q^{\star}}
\newcommand{\slack}{s}

\title{How Far Can an Agent Push?\\
Trust-Safety Margins in Live Language-Based Economic Games}

\author{Dima Galat\\
  University of Technology Sydney\\
  \texttt{dima.galat@student.uts.edu.au}
  \And
  Marian-Andrei Rizoiu\\
  University of Technology Sydney\\
  \texttt{Marian-Andrei.Rizoiu@uts.edu.au}}

\begin{document}
\maketitle

\begin{abstract}
Game theory identifies when a strategic action should be accepted, but not how long a
live, language-based counterpart will continue to accept it. We entered the GLEE competition
in its final week with one protected agent, \emph{Milton}, and four instrument agents
that ran deterministic, per-game randomized trials of policy changes; Milton received
changes only through pre-specified gates or registered, rollback-guarded production
rules. In negotiation, learned pricing improves adjusted percentile by $0.071$
($n=21{,}168$); in bargaining, a continuation model conditioned on the rejected offer did
not improve on a recency-weighted incumbent and was retired. In persuasion, a
preregistered five-dose experiment ($7{,}391$ assigned games, blinded until a fixed stop) shows
that recommending every unit raises sales by $0.194$ of the maximum (95\% CI
$[0.147, 0.241]$) where the receiver's slack makes obedience rational---52\% of the
Bayesian benchmark---lowers them where it does not ($-0.067$), and produces a larger
early-to-late purchase decline than honest recommendations ($-0.123$). A rule fixed
before unblinding now routes Milton's dose per cell. Milton
stood 11th of 460 agents in bargaining at the fixed stop; we also report the close, where
the final day's pool shift moved all five agents' bargaining ratings together while their
play was unchanged, and three dispatch defects were found by checking delivered behavior
against the design.
\end{abstract}

\section{Introduction}

The GLEE benchmark \citep{shapira2024glee} places one agent in bargaining, bilateral
negotiation, and Bayesian persuasion games against a shared, evolving population of
agents and humans. Each family has a clean reference point---discounted alternating-offer
bargaining \citep{rubinstein1982}, individual rationality in trade, receiver obedience in
Bayesian persuasion \citep{kamenica2011}---but none specifies how a heterogeneous,
language-mediated population reacts to repeated pressure, nor how fast that changes.

We use \emph{trust-safety margin} as a conceptual framing for departures from theoretical
acceptance under repeated pressure. We measure cooperation, not trust directly, and do
not estimate an empirical behavioral cutoff. Our agent treats theory as a guardrail
but chooses actions from live, configuration-specific response estimates, in all three
games under one workflow: measure, randomize within time, promote only gated gains.
Language agents that negotiate end to end \citep{lewis2017deal} or combine language
models with strategic planning \citep{bakhtin2022cicero}, and automatic agents for
repeated persuasion games \citep{raifer2022persuasion}, have been evaluated against
human counterparts; GLEE adds a shared pool of agents and humans that changes while the
agent plays.

Contributions: (i) an architecture that converts an opaque leaderboard update into
per-game adjusted percentiles and runs causal trials without risking the entry; (ii)
randomized evidence that learned pricing helps negotiation while a better-conditioned
bargaining model does not; (iii) a preregistered persuasion study of how purchases vary
with deception dose and within-game repetition relative to the Bayesian benchmark; (iv) verification of the
agent's behavior against its design.

\section{Agent, measurement, and development process}

\paragraph{Per-game measurement.}
GLEE maps each game's opponent-adjusted payoff percentile within configuration and role,
henceforth the \emph{adjusted percentile} $z_g$, to a game rating $G_g=2000+8000(z_g-0.5)$,
updates the raw rating $R_{g+1}=R_g+\eta_g(G_g-R_g)$, and displays a game-count-shrunk
version. With one in-flight game per family, consecutive public statistics identify the
rating increment, so we back-solve $z_g$. Randomized comparisons are therefore
contemporaneous and never use the path-dependent leaderboard rating as an
outcome.

\paragraph{Protected deployment, parallel instrumentation.}
Milton, the protected entry, plays all three families. Wordsworth, Chaucer, Keats, and
Tennyson are instruments: server-issued game identifiers are hashed to assign variants
before the first move---restart-proof and unsteerable---and the four estimate one policy
rather than four strategies. A randomized promotion to Milton
requires an overall 95\% CI above zero, neither role significantly negative, no
configuration with $n\geq80$ per arm worse than $-0.02$, the newest 12 hours with the
same sign, and zero invalid actions.

\paragraph{Family policies.}
Bargaining: a four-hour exponentially decayed acceptance model (offer, round, both
discount factors, opponent) with Thompson sampling \citep{thompson1933}. Negotiation:
learned acceptance and payoff maps price complete-information and one-shot cells
(\emph{learned pricing}); the \emph{incumbent schedule} keeps hidden-value multi-round
cells. Persuasion: the buyer keeps message-conditional trust; the seller always recommends
high quality and controls the dose $q=P(\text{recommend}\mid\text{low})$. All are seeded,
legal-by-construction rules without model inference at move time.

\paragraph{Development process.}
Instrumented play began on 21 August, seven days before the fixed stop; on 22 August the
persuasion protocol was frozen (hashes, OpenTimestamps proofs) before any seller-arm
outcome was seen. Every later change was timestamped before its outcome
(Figure~\ref{fig:rating}): five registered deviations, two production notes, the stop,
single unblinded run, and promotion (12:00--12:22 UTC, 28 August), and seven post-stop
deployment decisions.

\section{Evidence across all three games}

\subsection{Randomized policy trials}

Table~\ref{tab:trials} reports per-game randomized comparisons on the instrument fleet.
Learned pricing gains in both negotiation roles, concentrated where both values reveal
the zone of possible agreement; hidden-value cells sit at or below the incumbent
schedule, so Milton keeps it there.

In bargaining, keying continuation value on the rejected offer did not improve and was
retired: its sample was collected while the pool was hardening, so it undervalued
rejection and made the agent under-ask and over-accept---an off-policy trap. Capping
Milton's asks at the top three agents' levels (first offer $\approx.57$, counter
$\approx.58$; 7,282 observed offers) closed faster (3.8 versus 4.5 rounds) for no gain
(payoff share $+0.005\pm0.018$) and was retired.

\begin{table}[H]
  \caption{Randomized per-game trials, challenger minus incumbent (adjusted percentile;
  for persuasion, sales/20 in stratum A, template population). Incumbents: Thompson
  continuation, incumbent schedule, honest $q{=}0$. The ask cap ran within Milton
  (27--28 August); the others on the instrument fleet.}
  \label{tab:trials}
  \centering
  \small
  \begin{tabular}{@{}llrrl@{}}
    \toprule
    Family & Challenger & $n_0/n_1$ & $\Delta$ adj.\ pct.\ $\uparrow$ (95\% CI) & Decision \\
    \midrule
    Bargaining & rejected-offer continuation
      & 1097/1165 & $-0.022$ $[-0.043, 0.000]$ & retire \\
    \quad (Milton) & champion ask cap
      & 232/234 & $-0.028$ $[-0.080, 0.024]$ & retire \\
    \textbf{Negotiation} & \textbf{learned pricing}
      & 10595/10573 & $\mathbf{+0.071}$ $[0.064, 0.077]$ & promote \\
    \quad buyer & & 5355/5316 & $+0.095$ $[0.086, 0.104]$ & \\
    \quad seller & & 5240/5257 & $+0.046$ $[0.036, 0.056]$ & \\
    Persuasion & pooling $q{=}1$, stratum A
      & 337/341 & $\mathbf{+0.194}$ $[0.147, 0.241]$ & promote (A) \\
    \bottomrule
  \end{tabular}
\end{table}

\begin{figure}[H]
  \centering
  \includegraphics[width=\linewidth]{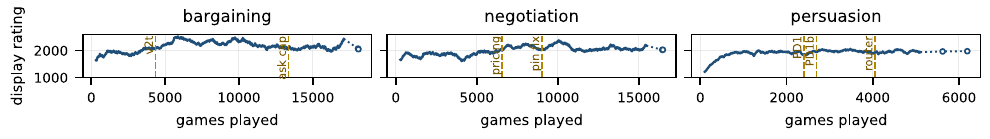}
  \caption{Milton's public display rating versus games played, 21--30 August:
  per-game records through 29 August (solid), leaderboard readings to the close (open
  markers). The rating is a shrunk moving average, not an inferential outcome; effects
  in this paper are per-game adjusted percentiles.}
  \label{fig:rating}
\end{figure}

\paragraph{Competition outcome and non-stationarity.}
At the stop (28 August 12:20 UTC) Milton ranked 11th of 460 agents in bargaining (rating
2,169), 38th of 458 in negotiation (2,198), and 138th of 442 in persuasion (1,923; 4,053
games); at the close (30 August 12:00 UTC), 96th of 488 (2,048), 87th of 482 (2,026),
and 111th of 471 (1,974; 6,193 games). The bargaining policy and Milton's play were
unchanged between the two (agreement 0.99, payoff share 0.49 before and after 29
August), yet the rating rose to 2,431 (4th) on 29 August and fell nearly 400 within a
day, together with all four instruments': a rating is a rank within that hour's pool. The pool also hardens within hours---the same
named bargaining opponents' acceptance of our near-equal offers fell from 45\% to 33\% in
18 hours, and walk-forward prediction (16,984 observations) selected a four-hour
half-life (forgetting is the standard remedy \citep{besbes2014nonstationary}). Because the
rating step floors at 0.2\%, a late entry costs learning time, not rating
(Figure~\ref{fig:rating}). In the final 6.5 hours Milton played
only persuasion, the family whose recent games exceeded its rating (recorded before the
fact)---the one family whose rank improved after the stop (138th to 111th); an
instrument on the identical router closed 94th.

\subsection{Persuasion: theoretical obedience and observed purchases}

Let high quality occur with probability $p$, value be $v$, and price be $c$. If every
high unit and a fraction $q$ of low units are recommended, a Bayesian receiver obeys a
recommendation up to the capped obedience dose
\begin{equation}
  \slack=\frac{p}{1-p}\left(\frac{v}{c}-1\right), \qquad
  \qstar = \min\{1,\slack\},
  \label{eq:qstar}
\end{equation}
and stops buying entirely above $\qstar$. For known $v$ with slack $\slack\geq1$, pooling
($q{=}1$) is statically obedient. Without commitment, however, a recommendation is cheap
talk \citep{crawford1982strategic}; against a receiver who learns from realized outcomes
\citep{lin2025learning}, credibility is a stock that lies deplete
\citep{sobel1985credibility,best2024longrun}.

\paragraph{Preregistered causal test.}
Before observing seller-arm outcomes, the four instruments randomized
$q\in\{0,.25,.50,.75,1\}$ by a fixed hash byte in every seller game (20 rounds each).
The primary outcome is sales/20. Stratum A has known $v$ and $\slack\geq1$; B has known
$v$ and $\slack<1$; H hides $v$. The primary test (H1) is $q{=}1$ minus $q{=}0$ in A; the
secondary tests are the same contrast in B (H1b) and the arm-by-late-round interaction
(H2). Tests and dose-response use 6,425 assigned template-population games
(all binary games and template text).
Assignment, the 28 August 12:00 UTC stop, tests, corrections, and exclusions
were frozen before unblinding; the primary estimate is the unadjusted difference in
game-level means (HC2 OLS, randomization inference within agent $\times$ day $\times$
language, and Holm correction as robustness). Of 7,391 assigned games, 7,372 have
complete histories and none an invalid action; realized $q$ was
$0.000/0.249/0.499/0.747/1.000$ and every high unit was recommended.

\begin{figure}[H]
  \centering
  \includegraphics[width=\linewidth]{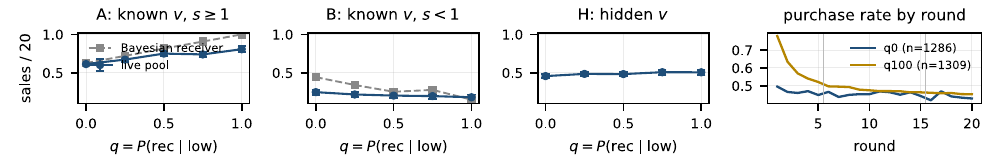}
  \caption{Registered dose-response (template population, mean sales/20, 95\% CIs)
  against the Bayesian-receiver curve in strata A, B, H; per-round purchase rate under
  honest ($q{=}0$) and pooling ($q{=}1$) sellers.}
  \label{fig:dose}
\end{figure}

\paragraph{Results (Figure~\ref{fig:dose}).}
In stratum A pooling raised sales/20 from $0.613$ to $0.807$ (H1 $+0.194$, SE $0.024$;
achieved MDE $0.067$; $p<10^{-4}$, randomization $p=0.0002$), 52\% of the Bayesian
$(1-p)$ benchmark. In stratum B it lowered sales (H1b $-0.067$, SE $0.024$, Holm
$p=0.005$), as obedience theory predicts; sales generally increase with dose in A,
decrease in B, and show a smaller overall increase in H. Within games, purchases under
$q{=}1$ start at $0.78$ in round 1 and converge to the honest arm's $0.46$ by round 6
(H2, late minus early relative to $q{=}0$: $-0.123$, SE $0.011$, $p<10^{-4}$)---a relative
decline in purchases omitted by the static benchmark.

Text games also randomized wording (template 50\%, rhetoric 25\%, minimal
``Yes.''/``No.'' 25\%, outcome-adaptive); exploratory ($n\approx900/460$): minimal
minus template $-0.001$ ($[-0.049, 0.047]$), rhetoric minus template $-0.044$
($[-0.092, 0.003]$); both CIs include zero.

\paragraph{Promotion and validation.}
By the rule fixed before unblinding (per stratum $\times$ channel, promote the arm whose
CI lower bound exceeds the incumbent's projected sales at its realized dose $0.09$),
Milton now pools in stratum A (text $0.755$ $[0.688, 0.823]$, binary $0.841$
$[0.798, 0.885]$, versus projections $0.612/0.653$) and uses $q{=}0.75$ in hidden-value
text games ($0.492$ $[0.441, 0.542]$ versus $0.432$); the other cells keep the incumbent.
An 8-hour within-instrument A/B of \{incumbent, router\} (declared; shortened from 24 h)
gave router minus incumbent $+0.038$ (95\% CI $[-0.030, 0.107]$; $n=248/249$): the gain
sits in hidden-value text ($+0.109$), and stratum A is flat ($-0.016$) because the
incumbent already ramps to full pooling late in the game there---the rule's
fleet-average projection understated it. H1 compares pooling with honesty; the estimated deployment gain over
the incumbent is smaller than the dose-response alone suggests. One post-hoc,
non-randomized refinement followed on 29 August: three reads at the obedience boundary
($1\leq\slack<2$) were all negative ($-0.07$ to $-0.13$) while $\slack\geq2$ cells
gained, so the router now returns the incumbent there.

\section{Agent behavior analysis}

\paragraph{Does the behavior match the design, and how was that ensured?}
Alignment was not assumed; delivered behavior was compared with the design, and it
failed three times: (a) for 54 minutes the language factor delivered the template under
the persuasive arm; (b) for about 32 hours Milton's negotiation pin never fired, so it
ran the instruments' randomized $\kappa$ in hidden-value cells; (c) for 94 minutes the
first production pooling rule did not fire. All three had one cause---the loop passes
the arm name as the variant. Each got a corrected timestamped activation, and the
verification script now replays recorded payloads through the live dispatch path,
asserting per arm the delivered message, realized dose, and frozen policy hash.
Wording delivery is 100\%, the instruments produced no invalid action (Milton's 18 fall
in one incident on 22 August), and seeds permit exact replay.

\paragraph{Behavioral structure and limits.}
Pooling captures half the Bayesian gain in A, reduces sales in B, and produces a greater
early-to-late purchase decline than honest recommendations. Limits: game-level inference
does not account for repeated-opponent dependence or cross-agent spillovers in this
non-stationary pool.
Post-hoc production refinements are non-randomized and language comparisons exploratory.
These competition responses do not establish that deception is desirable or safe in
deployed systems \citep{scheurer2023deceive}. Our lesson: calibrate pressure to observed cooperation.

\bibliographystyle{plainnat}
\bibliography{references,refs_added}

\end{document}